\documentclass[a4paper]{article}
\pdfoutput=1

\usepackage[utf8]{inputenc}
\usepackage[nocompress]{cite}
\usepackage{fullpage}
\usepackage{hyperref}
\usepackage{amsmath, amssymb}
\usepackage{graphicx}
\usepackage{amsthm} 
\usepackage{xcolor}
\usepackage{cleveref}
\usepackage{fontawesome5}

\newtheorem{corollary}{Corollary}
\newtheorem{theorem}{Theorem}
\newtheorem{lemma}{Lemma}

\def\orcidsymbol{\includegraphics[height=9pt]{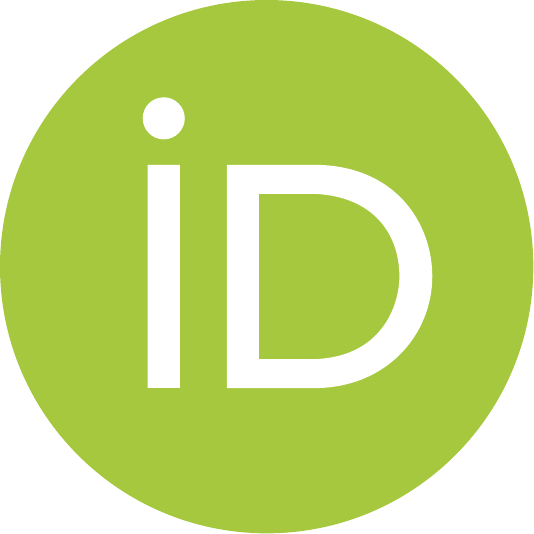}}
\def\mailsymbol{\textcolor{lightgray}{\fontsize{9}{12}\sffamily\bfseries \faIcon[regular]{envelope}}}

\date{}

\newcommand{\lcp}{\ensuremath\mathsf{lcp}}
\def\patterns{\mathcal{P}}

\def\nodeString#1{s_{#1}} % concatenated edge labels from root to node

\newcommand{\tmr}{truncate match reporting}
\newcommand{\catr}{colored ancestor threshold reporting}

\newcommand{\TRLE}{\ensuremath{T_{\textrm{RLE}}}}

\begin{document}

\title{Compressed Dictionary Matching on Run-Length Encoded Strings}

\author{Philip Bille~\href{mailto:phbi@dtu.dk}{\mailsymbol}\href{https://orcid.org/0000-0002-1120-5154}{\orcidsymbol}$^{1}$ \and
Inge Li G{\o}rtz~\href{mailto:inge@dtu.dk}{\mailsymbol}\href{https://orcid.org/0000-0002-8322-4952}{\orcidsymbol}$^{1}$ \and
Simon J.\ Puglisi~\href{mailto:simon.puglisi@helsinki.fi}{\mailsymbol}\href{https://orcid.org/0000-0001-7668-7636}{\orcidsymbol}$^{2}$ \and
Simon Rumle Tarnow~\href{mailto:sruta@dtu.dk}{\mailsymbol}\href{https://orcid.org/0009-0002-4293-6475}{\orcidsymbol}$^{1}$\\[0.5em]
{\small\begin{minipage}{\linewidth}\begin{center}
\begin{tabular}{c}
$^{1}$DTU Compute, Technical University of Denmark, Denmark\\
$^{2}$Department of Computer Science, University of Helsinki, Finland
\end{tabular}
\end{center}\end{minipage}}
}

\maketitle

\begin{abstract}
Given a set of pattern strings $\patterns=\{P_1, P_2,\ldots P_k\}$ and a text string $S$, the classic dictionary matching problem is to report all occurrences of each pattern in $S$. 
We study the dictionary problem in the compressed setting, where the pattern strings and the text string are compressed using run-length encoding, and the goal is to solve the problem without decompression and achieve efficient time and space in the size of the compressed strings. Let $m$ and $n$  be the total length of the patterns $\patterns$ and the length of the text string $S$, respectively, and let $\overline{m}$ and $\overline{n}$ be the total number of runs in the run-length encoding of the patterns in $\patterns$ and $S$, respectively. Our main result is an algorithm that achieves $O( (\overline{m} + \overline{n})\log \log m + \mathrm{occ})$ expected time, and $O(\overline{m})$ space, where $\mathrm{occ}$ is the total number of occurrences of patterns in $S$. This is the first non-trivial solution to the problem. Since any solution must read the input, our time bound is optimal within an $\log \log m$ factor.  We introduce several new techniques to achieve our bounds, including a new compressed representation of the classic Aho-Corasick automaton and a new efficient string index that supports fast queries in run-length encoded strings. 
\end{abstract}

\section{Introduction}
Given a set of pattern strings $\patterns=\{P_1, P_2,\ldots P_k\}$ and a string $S$, the \emph{dictionary matching problem} (also called the \emph{multi-string matching problem}) is to report all occurrences of each pattern in $S$. Dictionary matching is a classic and extensively studied problem in combinatorial pattern matching~\cite{AC75,DBLP:conf/icalp/Commentz-Walter79, DBLP:conf/focs/AmirF91,DBLP:journals/jcss/AmirFGBP94,DBLP:journals/iandc/AmirFIPS95,DBLP:journals/iandc/FerraginaL98,B10,B12,HKSTV13,DBLP:conf/cpm/AmirLPS14,INIBT15,DBLP:conf/esa/CliffordFPSS15,DBLP:conf/esa/0001GGK15,DBLP:journals/tcs/AmirLPS15,DBLP:conf/swat/0002HS16,DBLP:conf/cpm/KopelowitzPR16,DBLP:journals/mscs/AtharBBGILP17,DBLP:conf/esa/GolanP17,DBLP:journals/algorithmica/AmirKLPPS19} with the first efficient solution due to Aho and Corasick~\cite{AC75} from the 1970'ties. Dictionary matching is also a key component in several other algorithms for combinatorial pattern matching, see e.g.~\cite{Navarro2001a, CGL2004, ALP2004, CL1994, BI2016, BT2010, BGVK2012, ABF97, FGGK2015}. 

A \emph{run} in a string $S$ is a maximal substring of identical characters denoted $\alpha^x$, where $\alpha$ is the character of the run and $x$ is the length of the run. The \emph{run-length encoding} (RLE) of $S$ is obtained by replacing each run $\alpha^x$ in $S$ by the pair $(\alpha,x)$. For example, the run-length encoding of $aaaabbbaaaccbaa$ is $(a,4),(b,3),(a,3),(c,2),(b,1),(a,2)$.

This paper focuses on \emph{compressed dictionary matching}, where $\mathcal{P}$ and $S$ are given in run-length encoded form. The goal is to solve the problem without decompression and achieve efficient time and space in terms of the total number of runs in $\mathcal{P}$ and $S$. Compressed dictionary matching has been studied for other compression schemes~\cite{HKSTV13, INIBT15, B10, KTSMA1998}, and run-length encoding has been studied for other compressed pattern matching problems~\cite{AB92, ABF97, ALS03, ALS99, S12, EHSAV08}. However, no non-trivial bounds are known for the combination of the two.

\subparagraph{Results}
We address the basic question of whether it is possible to solve compressed dictionary matching in near-linear time in the total number of runs in the input. We show the following main result.   

\begin{theorem}
\label{thm:rle_reporting}
Let $\mathcal{P} = \{P_1, \ldots, P_k\}$ be a set of $k$ strings of total length $m$ and let $S$ be a string of length $n$. Given the RLE representation of $\mathcal{P}$ and $S$ consisting of $\overline{m}$ and $\overline{n}$ runs, respectively, we can solve the dictionary matching problem in $O((\overline{m} + \overline{n})\log{\log{m}} + \mathrm{occ})$ expected time and $O(\overline{m})$ space, where $\mathrm{occ}$ is the total number of occurrences of $\mathcal{P}$ in $S$. 
\end{theorem}
Theorem~\ref{thm:rle_reporting} assumes a standard word-RAM model of computation with logarithmic word length, and space is the number of words used by the algorithm, excluding the input strings, which are assumed to be read-only. 

This is the first non-trivial algorithm for compressed dictionary matching on run-length encoded strings. Since any solution must read the input the time bound of Theorem~\ref{thm:rle_reporting} is optimal within a $\log \log m$ factor. 

Furthermore, we demonstrate that we can achieve nearly the same complexity deterministically, as shown in the following result.
\begin{theorem}
\label{thm:rle_reporting_determ}
Let $\mathcal{P} = \{P_1, \ldots, P_k\}$ be a set of $k$ strings of total length $m$ and let $S$ be a string of length $n$. Given the RLE representation of $\mathcal{P}$ and $S$ consisting of $\overline{m}$ and $\overline{n}$ runs, respectively, we can deterministically solve the dictionary matching problem in $O((\overline{m} + \overline{n})\log{\log{(m + \overline{n})}} + \mathrm{occ})$ time and $O(\overline{n} + \overline{m})$ space, where $\mathrm{occ}$ is the total number of occurrences of $\mathcal{P}$ in $S$. 
\end{theorem}

The additional $\log{\log{(m + \overline{n})}}$ factor arise from a reduction of the alphabet that we achieve through sorting the runs in the patterns $\mathcal{P}$ and string $S$. Since we sort $S$ we also require an additional $O(\overline{n})$ space compared to the randomized result \cref{thm:rle_reporting}.

\subparagraph{Techniques}
Our starting point is the classic Aho-Corasick algorithm for dictionary matching~\cite{AC75} that generalizes the Knuth-Morris-Pratt algorithm~\cite{KMP77} for single string matching to multiple strings. Given a set of pattern strings $\patterns = \{P_1,\ldots,P_k\}$ of total length $m$
the \emph{Aho-Corasick automaton} (AC automaton) for $\patterns$ consists of the trie of the patterns in $\patterns$. Hence, any path from the trie's root to a node $v$ corresponds to a prefix of a pattern $P\in\patterns$. For each node $v$, a special \emph{failure link} points to the node corresponding to the longest prefix matching a proper suffix of the string identified by $v$ and an \emph{output link} that points to the node corresponding to the longest pattern that matches a suffix of the string identified by $v$. 

To solve dictionary matching, we read $S$ one character at a time and traverse the AC automaton. At each step, we maintain the node corresponding to the longest suffix of the current prefix of $S$. If we cannot match a character, we recursively follow failure pointers (without reading further in $S$). If we reach a node with an output link, we output the corresponding pattern and recursively follow output links to report all other patterns matching at the current position in $S$. If we implement the trie using perfect hashing~\cite{DBLP:conf/focs/FredmanKS82}, we can perform the top-down traversal in constant time per character. Since failure links always point to a node of strictly smaller depth in the trie, it follows that we can charge the cost of traversing these to reading characters in $S$, and thus, the total time for traversing failure links is  $O(n)$. Each traversal of an output link results in a reported occurrence, and thus, the total time for traversing output links is $O(\mathrm{occ})$. In total, this leads to a solution to dictionary matching that uses $O(m)$ space and $O(n + m + \mathrm{occ})$ expected time.

At a high level, our main result in Theorem~\ref{thm:rle_reporting} can viewed as a compressed implementation of the AC-automaton, that implements dictionary matching  $O((\overline{m} + \overline{n})\log{\log{m}} + \mathrm{occ})$ expected time and $O(\overline{m})$ space. Compared to the uncompressed AC-automaton, we achieve the same bound in the compressed input except for a $\log \log m$ factor in the time complexity.

To implement the AC automaton in $O(\overline{m})$ space, we introduce the \emph{run-length encoded trie} $\TRLE$ of $\patterns$, where each run $\alpha^x$ in a pattern $P\in \patterns$ is encoded as a single letter `$\alpha^x$' and show how to simulate the action of the Aho-Corasick algorithm on $\TRLE$ processing one run of $S$ at a time. The key challenge is that the nodes in the AC automaton are not explicitly represented in $\TRLE$, and thus, we cannot directly implement the failure and output links in $\TRLE$. 

Naively, we can simulate the failure links of the AC automaton directly on $\TRLE$. However, as in the Aho-Corasick algorithm, this leads to a solution traversing $\Omega(n)$ failure links, which we cannot afford. Instead, we show how to efficiently construct a data structure to group and process nodes simultaneously, achieving $O(\overline{n}\log \log m)$ total processing time.

Similarly, we can simulate the output links by grouping them at the explicit nodes, but even on a simple instance such as $\patterns=\{a, a^{m-1}\}$ this would result in $\Omega(m)$ output links in total which we also cannot afford.
Alternatively, we can store a single output link for each explicit node as in the AC automaton. However, the occurrences we need to report are not only patterns that are suffixes of the string seen until now, but also patterns ending inside the last processed run. Not all occurrences ending in the last  run are substrings of each other, and thus, saving only the longest is not sufficient.
Instead, we reduce the problem of reporting all occurrences that end in a specific run of $S$ to a new data structure problem called \emph{\tmr} problem. To solve this problem, we reduce it to a problem of independent interest on colored weighted trees called \emph{\catr}. We present an efficient solution to the \tmr\ problem by combining several existing tree data structures in a novel way, ultimately leading to the final solution. 

Along the way, we also develop a simple and general technique for \emph{compressed sorting} of run-length encoded strings of independent interest, which we need to preprocess our pattern strings efficiently. 

Finally, we show that we can achieve almost the same result deterministically.
The key challenge here is to answer dictionary queries deterministically while avoiding dependency on the size of the alphabet $\sigma$.
We show that we can deterministically reduce the size of the alphabet $\sigma$ to $O(\min(\overline{n}, \overline{m}) + 1)$ in
$O((\overline{n}+\overline{m})\log{\log{(\overline{n} + \overline{m}})})$ time.
We can then use predecessor search to answer dictionary queries without affecting the overall time complexity.

\subparagraph{Outline}
In \cref{sec:sorting_and_tries}, we show how to efficiently sort run-length encoded strings and subsequently efficiently construct the corresponding compact trie.
In \cref{sec:reporting} we introduce the \emph{\tmr} and \emph{\catr} problem, our efficient solution to the \emph{\catr} problem, and our reduction from the \emph{\tmr} problem to the \emph{\catr} problem.
In \cref{sec:static:compressed}, we first give a simple and inefficient version of our algorithm, that focuses on navigation the run-length encoded trie.
In \cref{sec:full_structure}, we extend and improve the simple version of our algorithm to achieve our main result. 
Finally, in \cref{sec:determ}, we adapt our structures to become deterministic.

\section{Preliminaries}
We use the following well-known basic data structure primitives. Let $X$ be a set of $n$ integers from a universe of size $u$. Given an integer $x$, a \emph{membership query} determines if $x \in X$. A \emph{predecessor query}, return the largest $y \in X$ such that $y \leq x$. We can support membership queries in $O(n)$ space, $O(n)$ expected preprocessing time, and constant query time using the well-known perfect hashing construction of Fredman, Koml\'{o}s, and Szemer\'{e}di~\cite{DBLP:conf/focs/FredmanKS82}. By combining the well-known $y$-fast trie of Willard~\cite{DBLP:journals/ipl/Willard83} with perfect hashing, we can support predecessor queries with the following bounds. 

\begin{lemma}\label{lm:predecessor}
Given a set of $n$ integers from a universe of size $u$, we can support predecessor queries in $O(n)$ space, $O(n \log \log n)$ expected preprocessing time, and $O(\log \log u)$ query time. 
\end{lemma}

\section{Sorting Run-Length Encoded Strings and Constructing Compact Tries}
\label{sec:sorting_and_tries}
Let $\mathcal{P} = \{P_1, \ldots, P_k\}$ be a set of $k$ strings of total length $m$ from an alphabet of size $\sigma$. Furthermore, let $\overline{\mathcal{P}} = \{\overline{P_1}, \ldots, \overline{P_k}\}$ be the RLE representation of $\mathcal{P}$ consisting of $\overline{m}$ runs. In this section, we show that given $\overline{\mathcal{P}}$ we sort the corresponding (uncompressed) set of strings in $\mathcal{P}$ and construct the compact trie of them in expected $O(\overline{m} + k \log \log k)$ time and $O(\overline{m})$ space. We use this compact trie construction to efficiently preprocess our patterns in our compressed dictionary matching algorithm in the following sections. 

We first demonstrate that a compact trie can be efficiently constructed for a set of strings, provided the strings are given in sorted order.
\begin{lemma}\label{lm:rle_compact_trie}
Given a set $\mathcal{P}= \{P_1, \ldots, P_k\}$ of $k$ strings in sorted order of total length $m$, 
we can construct the compact trie $T$ of $\mathcal{P}$ in $O(m)$ time and space.
\end{lemma}

\begin{proof}
Our algorithm proceeds as follows: 
\subparagraph{Step 1: Compute Longest Common Prefixes}
We compute the longest common prefixes $\ell_1, \ldots, \ell_{k-1}$, where $\ell_i = \lcp(P_i, P_{i+1})$, of each consecutive pairs of strings in $\mathcal{P}$. To do so, we scan each pair of strings $P_i$ and $P_{i+1}$ from left to right to find the longest common prefix.

\subparagraph{Step 2: Construct the Compact Trie}
To construct the compact trie $T$, we add the strings one at a time from left to right to an initially empty trie. We maintain the string depth of each node during the construction. Suppose we have constructed the compact trie $T_i$ of the strings $\{P_1, \ldots, P_i\}$ and consider the leftmost path $p$ in $T_i$ corresponding to $P_{i}$. Let $P_{i+1}'$ denote the suffix of $P_{i+1}$ not shared with $P_i$. We add the string $P_{i+1}$ to $T_i$ as follows. If $\ell_i = |P_i|$, we extend the path $p$ with a new edge representing $P_{i+1}'$. Otherwise, we traverse $p$ bottom up to find the location at string depth $\ell_i$ to attach a new edge representing $P_{i+1}'$. Note that this location is either an existing node in $T$ or we need to split an edge and add a new node. At the end, we return the trie $T = T_k$. 
\medskip

Since the strings are sorted, step 2 inductively constructs the compact trie $T_i$, for $i = 1,\ldots, k$ of the $i$ first strings. Thus, $T = T_k$ is the compact trie of $\mathcal{P}$. Step 1 uses $O(m)$ time.
For step 2, we bound the total number of edges in the bottom-up traversal of the rightmost path. Consider traversing a rightmost path $p$ in $T_i$ while adding $P_{i+1}$ and let $v$ be the (existing or newly created) node of string depth $\ell_i$ on $p$ where we attach a new child edge representing $P_{i+1}'$. The edges on $p$ below $v$ are never traversed again and are part of the final trie $T$. Hence, the total time to traverse them is at most $O(k) = O(\overline{m})$. The remaining parts of step 2 take $O(m)$ time, and hence, the total time is $O(m)$. We use $O(k + m) = O(m)$ space in total. 
\end{proof}

We now show the following simple reduction to standard (uncompressed) string sorting. Let $t(k, m, \sigma)$ and $s(k, m, \sigma)$ denote the time and space, respectively, to sort $k$ strings of total length $m$ from an alphabet of size $\sigma$.

\begin{lemma}\label{lm:rle_sort_blackbox}
    Let $\mathcal{P} = \{P_1, \ldots, P_k\}$ be a set of $k$ strings of total length $m$ from an alphabet of size $\sigma$. Given the RLE representation $\overline{\mathcal{P}}=\{\overline{P_1}, \ldots, \overline{P_k}\}$ of $\mathcal{P}$ consisting of $\overline{m}$ runs, we can sort $\mathcal{P}$ in $O(t(k, \overline{m}, \sigma \cdot m) + \overline{m})$ time and $O(s(k, \overline{m}, \sigma \cdot m) + \overline{m})$ space. 
\end{lemma}

\begin{proof}
To sort the strings $\mathcal{P}$, we construct the set of strings $\tilde{\mathcal{P}} = \{\tilde{P}_1, \ldots, \tilde{P}_k\}$ such that $\tilde{P}_i=(\alpha_1, x_1)(\alpha_2, x_2)\cdots$ is the sequence of the runs in $\overline{P_i}=\alpha_1^{x_1}\alpha_2^{x_2}\cdots$, represented as pairs of character and length.
We say that a pair $(\alpha, x)$ is smaller than $(\beta, y)$ if $\alpha < \beta$ or $x < y$.
Assume that we have the compact trie $\tilde{T}$ of $\tilde{\mathcal{P}}$, where the edges are lexicographically sorted by their edge labels.
We will show that we can construct the compact trie $T$ of $\mathcal{P}$ from $\tilde{T}$ in $O(\overline{m})$ time and maintain the ordering of the edges.
Observe that for a node in $v\in T$ the ordering of edge labels of the children of $v$ are equivalent to the ordering achieved by encoding them as pairs of character and length, since the first character of each edge would differ.
Let $w$ be an internal node in $T$ such that there is no node $\tilde{w} \in \tilde{T}$ where $s_w = s_{\tilde{w}}$.
Since $w$ is an internal node in $T$, two patterns $P_i$ and $P_j$ exist such that the longest common prefix of $P_i$ and $P_j$ is $s_w$.
Since $w$ does not have a corresponding node in $\tilde{T}$, then there must exists a node $\tilde{w}$ in $\tilde{T}$ such that $s_w=s_{\tilde{w}}\beta^y$ with at least two children with edge labels $(\beta, y)Y$ and $(\beta, z)Z$ where $y < z$.
We ensure that any node $w\in T$ has a corresponding node in $\tilde{T}$ as follows.
Starting at the root of $\tilde{T}$, we do the following at each node $v$.
Let $v_1, v_2, \ldots, v_d$ be the children of $v$ lexicographically ordered by their edge labels $(\beta_1, y_1)Y_1, (\beta_2, y_2)Y_2, \ldots, (\beta_d, y_d)Y_d$.
Starting at $i=d$ we do the following.
\begin{itemize}
    \item If $\beta_{i-1} = \beta_i$ and $Y_{i-1}$ is not the empty string, we insert a node $w$ on the edge to $v_{i-1}$ and make $v_{i}$ the child of $w$.
    We then set the edge label of edge $(v,w)$, $(w, v_{i-1})$, and $(w, v_i)$ to $(\beta_i, y_{i-1})$, $Y_{i-1}$, and $(\beta_i, y_{i} - y_{i-1})Y_i$, respectively. 
    \item If $\beta_{i-1} = \beta_i$ and $Y_{i-1}$ is the empty string, we make $v_i$ the child of $v_{i-1}$ and set the edge label of $(v_{i-1}, v_i)$ to $(\beta_i, y_{i} - y_{i-1})Y_i$. 
\end{itemize}
Note that if $Y_{i-1}$ is empty, then no child of $v_{i-1}$ will have $(\beta_i, z)$ as the first character on their edge label, since it could then be encoded as $(\beta_i, y_{i-1} + z)$. 
Hence, the parent of $v_i$ cannot change again.
To maintain the order of the children of $v_i$ and $w$, we insert new edges using insertion sort.
When $w$ is created, it takes $v_{i-1}$'s place in the ordering of the children of $v$.
We then decrement $i$ until $i=1$ and then recurse on the remaining children of $v$.
Finally, we traverse $\tilde{T}$ and remove non-root vertices with only a single child.
Since the children already appear in sorted order, we can recover the sorted order of $\mathcal{P}$ from the constructed compact trie.
We use $O(t(k, \overline{m}, \sigma\cdot m) + \overline{m})$ time and $O(s(k, \overline{m}, \sigma\cdot m) + \overline{m})$ space to sort $\tilde{\mathcal{P}}$.
We can construct $\tilde{T}$ from the sorted sequence of $\tilde{\mathcal{P}}$ in $O(\overline{m})$ time due to \cref{lm:rle_compact_trie}.
We insert at most $O(k)$ nodes in $\tilde{T}$, and we apply insertion sort $O(1)$ time at each node. Hence, we use $O(\overline{m})$ time doing the insertion sort.
In summary, we have shown \cref{lm:rle_sort_blackbox}.
\end{proof}

Andersson and Nilsson~\cite{DBLP:conf/focs/AnderssonN94} showed how to sort strings in $t(k, m, \sigma) = O(m + k\log{\log{k}})$ expected time and $s(k, m, \sigma) = O(m)$ space. We obtain the following result by plugging this into \cref{lm:rle_sort_blackbox}. 

\begin{corollary}\label{cor:rle_sort}
    Let $\mathcal{P} = \{P_1, \ldots, P_k\}$ be a set of $k$ strings of total length $m$ from an alphabet of size $\sigma$. Given the RLE representation $\overline{\mathcal{P}}=\{\overline{P_1}, \ldots, \overline{P_k}\}$ of $\mathcal{P}$ consisting of $\overline{m}$ runs, we can sort $\mathcal{P}$ in  $O(\overline{m} + k\log \log k)$ expected time and $O(\overline{m})$ space. 
\end{corollary}

Using \cref{cor:rle_sort} and \cref{lm:rle_sort_blackbox} we can efficiently construct the compact trie of $\mathcal{P}$ from $\overline{\mathcal{P}}$.

\begin{lemma}\label{cor:rle_compact_trie}
    Let $\mathcal{P} = \{P_1, \ldots, P_k\}$ be a set of $k$ strings of total length $m$ from an alphabet of size $\sigma$. Given the RLE representation $\overline{\mathcal{P}}=\{\overline{P_1}, \ldots, \overline{P_k}\}$ of $\mathcal{P}$ consisting of $\overline{m}$ runs, we can construct the compact trie of $\mathcal{P}$ in  $O(\overline{m} + k\log{\log{k}})$ expected time and $O(\overline{m})$ space.
\end{lemma}

\section{Truncated Matching}
\label{sec:reporting}
\begin{figure}[t]
    \centering
    \includegraphics[scale=0.8]{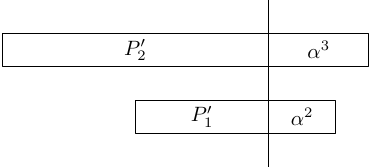}
    
    \caption{
    The strings $P_1$ and $P_2$ where $P_1'$ and $P_2'$ is the truncated string of $P_1$ and $P_2$, respectively.
    If $P_1'$ is a suffix of $P_2'$ then $P_1$ truncate matches $P_2$ since the length of the last run in $P_1$ is no longer than in $P_2$ and is the same character.
    }
    \label{fig:truncate}
\end{figure}

This section presents a compact data structure that efficiently supports a new type of pattern matching queries that we call \emph{truncated matching}.  We will need this result to implement output links in our algorithm efficiently. Given a string $P$, we define the \emph{truncated string} of $P$ to be the string $P'$ such that $P=P'\alpha^w$, where $\alpha^w$ is the last run in $P$. 
Let $P_1$ and $P_2$ be two strings of the form 
$P_1=P_1'\alpha_1^{w_1}$ and $P_2= P_2'\alpha_2^{w_2}$ where $\alpha_1^{w_1}$ and $\alpha_2^{w_2}$ are the last runs of $P_1$ and $P_2$, respectively. We say that $P_1$ \emph{truncate matches} $P_2$ if $P_1'$ is a suffix of $P_2'$, $\alpha_1=\alpha_2$, and $w_1\leq w_2$. (See Figure~\ref{fig:truncate}).

Let $\mathcal{P} = \{P_1, \ldots, P_k\}$ be a set of $k$ strings of the form $P_i=P_i'\alpha_i^{w_i}$, where $\alpha_i^{w_i}$ is the last run of $P_i$ for all $i$. The \emph{\tmr} problem is to construct a data structure such that given an index $1\leq i\leq k$, a character $\alpha$,  and an integer $w$, one can efficiently report all indices $j$, such that $P_j$ truncate matches~$P_i'\alpha^w$.

Our goal in this section is to give a data structure for the truncate match reporting problem that uses  $O(k)$ space, $O(\overline{m} + k\log \log k)$ expected preprocessing time, and supports queries in $O(\log \log k)$ time. 

\subsection{Colored Ancestor Threshold Reporting}
We first define a data structure problem on colored, weighted trees, called \emph{colored ancestor threshold reporting}, and present an efficient solution. In the next section, we reduce truncate match reporting to colored threshold reporting to obtain our result for truncate match reporting.  

Let $\mathcal{C}$ a set of \emph{colors}, and let $T$ be a rooted tree with $n$ nodes, where each node $v$ has a (possibly empty) set of colors $C_v \subseteq \mathcal{C}$ and a function $\pi_v:C_v \rightarrow \{1,\ldots,W\}$ that associates each color in $C_v$ with a \emph{weight}.
The \emph{colored ancestor threshold reporting problem} is to construct a data structure such that given a node $v$, a color $c$, and an integer $w$, one can efficiently report all ancestors $u$ of $v$, such that $c\in C_u$ and $\pi_u(c) \leq w$. 

We present a data structure for the colored ancestor threshold reporting problem that uses $O(n + \sum_{v\in T}|C_v|)$ space, $O(n + \sum_{v\in T}|C_v|)$ expected preprocessing time, and supports queries in $O(\log\log n +\text{occ})$ time, where $\text{occ}$ is the number of reported nodes.

\subparagraph{Data Structure}
Our data structure stores the following information. 
\begin{itemize}
\item The tree $T$ together with a \emph{first color ancestor} data structure that supports queries of the form: given a node $v$ and a color $c$ return the nearest (not necessarily proper) ancestor $u$ of $v$ such that $c\in C_u$.
\item Furthermore, for each color $c \in \mathcal{C}$ we store: 
\begin{itemize}
        \item An \emph{induced tree} $T_c$ of the nodes $v\in T$ that have color $c\in C_v$, maintaining the ancestor relationships from $T$. Each node $v \in T_c$ has a weight $w_v = \pi_v(c)$.
         \item A \emph{path minima} data structure on $T_c$ that supports queries of the form: given two nodes $u$ and $v$ return a node with minimum weight on the path between $u$ and $v$ (both included). 
        \item A \emph{level ancestor} data structure on $T_c$ that supports queries of the form: given a node $v$ and an integer $d$ return the ancestor of $v$ of depth $d$.
\end{itemize}
\item For each node $v\in T$: a dictionary associating each color in $C_v$ with the corresponding node in $T_c$. 
\end{itemize}
The total size of the induced trees is $O(\sum_{v\in T}|C_v|)$. Hence, we can construct these and the associated dictionaries at each node in a single traversal of $T$ using $O(n + \sum_{v\in T}|C_v|)$ space and $O(n + \sum_{v\in T}|C_v|)$ expected preprocessing time. We use standard linear space and preprocessing time and constant query time path minima and level-ancestor data structures on each of the induced trees~\cite{DBLP:journals/algorithmica/Chazelle87a, DBLP:conf/icalp/AlstrupH00, BerkmanV94, Dietz91}. We use a first color ancestor data structure, which uses linear space, expected linear preprocessing time, and $O(\log\log n)$ query time~\cite{DBLP:conf/wads/Dietz89a, DBLP:conf/soda/MuthukrishnanM96, DBLP:conf/esa/FerraginaM96, DBLP:conf/focs/AlstrupHR98}. In total, we use $O(n + \sum_{v\in T}|C_v|)$ space and $O(n + \sum_{v\in T}|C_v|)$ expected preprocessing time. 

\subparagraph{Query}
Consider a query $(u, c, w)$. We perform a first colored ancestor query $(u,c)$ in $T$ that returns a node $u'$. We then look up the node $v$ in $T_c$ corresponding to $u'$ and perform a path minima query between $v$ and the root of $T_c$. Let $x$ be the returned node. If $w_x > w$, we stop. Otherwise, we return the node $x$.
Finally, we recurse on the path from the root of $T_c$ to the parent of $x$ and the path from $v$ to the child $x'$ of $x$ on the path to $v$. To find $x'$, we use a level-ancestor query on $v$ with $d$ equal to the depth of $x$ plus one.

The first colored ancestor query takes $O(\log\log n)$ time. Since a path minima query takes constant time, each recursion step uses constant time. The number of recursive steps is $O(1+\text{occ})$, and hence, the total time is $O(\log\log n + \text{occ})$. In summary, we have shown the following result. 

\begin{lemma}\label{lm:catr}
Let $\mathcal{C}$ be a set of colors and let $T$ be a rooted tree with $n$ nodes, where each node $v$ has a (possibly empty) set of colors $C_v \subseteq \mathcal{C}$ and a weight function~$\pi_v:C_v \rightarrow \{1,\ldots,W\}$. We can construct a data structure for the \catr\ problem that uses $O(n + \sum_{v\in T}|C_v|)$ space, $O(n + \sum_{v\in T}|C_v|)$ expected preprocessing time, and supports queries in $O(\log\log n +\text{occ})$ time, where $\text{occ}$ is the number of reported nodes. 
\end{lemma}

\subsection{Truncate Match Reporting}
We now reduce truncate match reporting to colored ancestor threshold reporting. 

\subparagraph{Data Structure} For each string $P_i \in\patterns$, let $\alpha_i^{w_i}$ be the last run of $P_i$ and let $P'_i$ be the truncated string of $P_i$.  We construct the compact trie $T$ of the reversed truncated strings  $\overleftarrow{P'_1}, \overleftarrow{P'_2}, \ldots, \overleftarrow{P'_k}$. Each $\overleftarrow{P'_i}$ corresponds to a node in $T$. Note that several truncated strings can correspond to the same node if the original strings end in different runs. 
For each node $v\in T$, let $I_v$ be the set of string indices whose truncated strings correspond to node $v$.

\begin{figure}[t]
    \centering
    \includegraphics[scale=1]{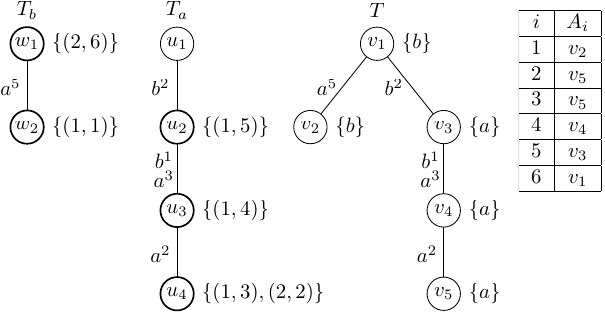}
    
    \caption{
    The structures of \cref{lm:catr} and \cref{lm:ctsr} for the strings $\{a^5b^1, a^5b^3a^2, a^5b^3a^1, a^3b^3a^1, b^2a^1, b^2\}$.
    To the right of each node $v\in T$ is the set of colors associated with $v$.
    To the right of each node $v$ in one of the trees $T_a$ and $T_b$ is the list $w_{v,\alpha}$ where $\alpha=a$ and $\alpha=b$, respectively.
    Finally $\pi_v(\alpha)$ is the first weight in $w_{v', \alpha}$ where $v'$ is the corresponding node in $T_\alpha$.
    }
    \label{fig:reporting}
\end{figure}

Our data structure consists of the following information.
\begin{itemize}
    \item The compact trie $T$.
    \item For each node $v$ in $T$ we store the following information:
    \begin{itemize}
        \item $C_v=\{\alpha_i \mid i \in I_v\}$. 
        \item A function $\pi_v: C_v \rightarrow \{1,\ldots, m\}$ where $\pi_v(\alpha) = \min_{i \in I_v}\{w_i \mid \alpha_i=\alpha\}$.
        \item For each $\alpha \in C_v$: A list $W_{v,\alpha}$ containing the set $\{(w_i, i) \mid i \in I_v \textrm{ and } \alpha_i = \alpha  \}$ sorted in increasing order of $w_i$.
        
    \end{itemize}
    \item A \catr\ data structure for $T$ with colors $C_v$ and weight functions $\pi_v$. 
    \item An array $A$ associating each string index with its corresponding node in the trie. 
\end{itemize}
See Figure~\ref{fig:reporting} for an example.

The compact trie $T$ uses space $O(k)$. The total size of the $I_v$ lists is $O(k)$ and thus $\sum_{v\in T}|C_v| = O(k)$.  It follows that the total size of the $W_{v,\alpha}$ lists is $O(k)$. The space of the \catr\ data structure is $O(k + \sum_{v\in T}|C_v|) = O(k)$. Thus the total size of the data structure is~$O(k)$.

We can construct the compact trie  $T$ in $O(\overline{m} + k\log{\log{k}})$ expected time  using \cref{cor:rle_compact_trie}. The preprocessing time of the \catr\ data structure is expected $O(k + \sum_{v\in T}|C_v|) = O(k)$.
Finally, we can sort all the $W_{v,\alpha}$ lists in $O(\sum_{v\in T}\sum_{\alpha\in C_v}|W_{v,\alpha}| \log\log{k}) = O(k\log\log{k})$ time \cite{DBLP:journals/jal/Han04}.
Thus, the preprocessing takes $O(\overline{m} + k\log{\log{k}})$ expected time.

\subparagraph{Queries} To answer a truncate match reporting query $(i, \alpha, w)$ we  perform a \catr\ query with $(A[i], \alpha, w)$. For each node $u$ returned by the query, we return all string indices in $W_{u, \alpha}$ with weight at most $w$ by doing a linear list scan and stop when the weight gets too big.

The \catr\ query takes $O(\log\log{k}+\mathrm{occ})$ time. We have at least one occurrence for each reported node, and the occurrences are found by a scan taking linear time in the number of reported indices. In total the query takes  $O(\log\log{k}+\text{occ})$~time.

In summary, we have shown the following result. 
\begin{lemma}\label{lm:ctsr}
Let $\mathcal{P} = \{P_1,\ldots,P_k\}$ be a set of $k$ strings. Given the RLE representation $\overline{\mathcal{P}} = \{\overline{P_1},\ldots,\overline{P_k}\}$ of $\mathcal{P}$
 consisting of $\overline{m}$ runs, 
we can construct a data structure for the \tmr\ problem using $O(k)$ space and expected $O(\overline{m} + k\log{\log{k}})$ preprocessing time, that can answer queries in time $O(\log\log{k}+\text{occ})$ where $\text{occ}$ is the number of reported occurrences.
\end{lemma}

\section{A Simple Solution}\label{sec:static:compressed}
In this section, we provide a simple solution that 
solves the compressed dictionary matching problem in $O(\overline{m})$ space and $O(\overline{n}\log{\log{k}} + \overline{m}\log{\log{m}} + n + m + \text{occ})$ expected time. In the next section, we show how to improve this to obtain the main result of \cref{thm:rle_reporting}.

\begin{figure}[t]
    \centering
    \includegraphics[scale=1]{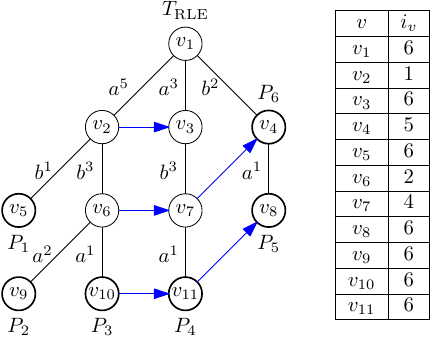}
    
    \caption{
    The structure of the simple solution for the 6 patterns $\{a^5b^1, a^5b^3a^2, a^5b^3a^1, a^3b^3a^1, b^2a^1, b^2\}$.
    Here the blue arrows indicate the failure link of a node.
    Failure links going to the root node have been left out for simplicity.
    Note that the patterns are the same as in \cref{fig:reporting}.
    }
    \label{fig:simple}
\end{figure}

Let $T_\text{RLE}$ be the \emph{run-length encoded trie} of $\patterns$, where each run $\alpha^x$ in a pattern $P\in \patterns$ is encoded as a single letter `$\alpha^x$'. See \cref{fig:simple} for an example.

The idea is to use $T_\text{RLE}$ to process $S$ one run at a time. At each step we maintain a position in $T_\text{RLE}$ such that the string of the root to leaf path is the longest suffix of the part of the text we have seen so far. To efficiently report matches we use the data structure for \tmr.

We will assume for now that the patterns $\patterns$ all have at least 2 runs.
Later we will show how to deal with patterns that consist of a single run.

\subparagraph{Data Structure} For a node $v$ in the trie $T_\text{RLE}$, let $\nodeString{v}$ denote the string obtained by concatenating the characters of the edges of the root-to-$v$ path where a run is not considered as a single letter.
We store the following for each node $v$ in $T_\text{RLE}$:
\begin{itemize}
  \item $D_v$: A dictionary of the children of $v$, with their edge labels as key.
  \item $F_v$: A failure link to the node $u$ in $T_\text{RLE}$ for which $\nodeString{u}$ is the longest proper suffix of $\nodeString{v}$. We define the failure link for the root to be itself.  
  \item $i_v$: The pattern index such that pattern $P'_{i_v}$ is the longest suffix of $\nodeString{v}$ among all truncated patterns.
  If no such pattern exists let $i_v = -1$.
\end{itemize}
We build the \tmr\ data structure for the set $\patterns$.
Finally, for each pattern $P_i=P_i'\alpha_i^{x_i}$ we store its length $|P_i|$ and the length $x_i$ of the last run of $P_i$.
See \cref{fig:simple} for an example.

The number of edges and nodes in $T_\text{RLE}$ is $O(\overline{m})$.
The dictionaries and failure links associated with $T_\text{RLE}$ use space proportional to the number of edges and vertices in $T_\text{RLE}$, hence $O(\overline{m})$ space.
By \cref{lm:ctsr} the \tmr\ data structure uses $O(k)$ space.
In total, we use $O(\overline{m} + k)= O(\overline{m})$ space.

\subparagraph{Query}
Given a run-length encoded string $S$ we report all locations of occurrences of the patterns $\patterns$ in $S$ as follows.
We process $S$ one run at a time.
Let $S'$ be the processed part of $S$.
We maintain the node $v$ in $T_\text{RLE}$ such that $\nodeString{v}$ is the longest suffix of $S'$.
Initially, $S'=\epsilon$ and $v$ is the root.
Let $\alpha^y$ be the next unprocessed run in $S$.
We proceed as follows:
\begin{itemize}
    \item[] {\bf Step 1: Report Occurrences.\;} Query the \tmr\ data structure with $(i_v, \alpha, y)$ unless $i_v=-1$.
    For each pattern $j$ returned by the query, report that pattern $j$ occurs at location $|S'| - |P_j| + x_j$.
    \item[] {\bf Step 2: Follow Failure Links.\;} While $v$ is not the root and $\alpha^y\notin D_v$ set $v=F_v$.
    Finally, if $\alpha^y\in D_v$ let $v=D_v[\alpha^y]$.
\end{itemize}

\subparagraph{Time Analysis}
At each node we visit when processing $S$, we use constant time to determine if we follow an edge or a failure link.
We traverse one edge per run in $S$ and thus traverse at most $O(\overline{n})$ edges. 
Since a failure link points to a proper suffix, the total number of failure links we traverse cannot exceed the number of characters we process throughout the algorithm.
Since we process $n$ characters we traverse at most $O(n)$ failure links.
Furthermore, we report from at most one node per run in $S$, and by \cref{lm:ctsr} we use $O(\overline{n}\log{\log{k}} + occ)$ time to report the locations of the matched patterns.
In total, we use $O(n + \overline{n}\log{\log{k}} + occ)$ time.

\subparagraph{Correctness}

 By induction on the iteration $t$, we will show that $\nodeString{v}$ is the longest suffix of $S'$ in $T_{\text{RLE}}$ and that we report all starting positions of occurrences of patterns $\patterns$ in $S'$.
 Initially, $t=0$, $S'=\epsilon$, and $v$ is the root of $T_\text{RLE}$ and thus $\nodeString{v}=\epsilon$.
 
 For the induction step assume that at the beginning of iteration $t$, $\nodeString{v}$ is the longest suffix of $S'$ in $T_\text{RLE}$.
 Let $\alpha^y$ be the next unprocessed run in $S$ and let $X\subseteq \mathcal{P}$ be the subset of patterns that has an occurrence that ends in the unprocessed run $\alpha^y$.
 Note that any pattern $P\in X$ has one occurrence that ends in the unprocessed run $\alpha^y$ since $P$ consists of at least two runs.

 We first argue that we correctly report all occurrences. It follows immediately from the definition of truncate match that $X$ consists of all the patterns $P\in\patterns$ that truncate matches $S'\alpha^y$. 
 Let $P_i'$ be the longest suffix of $S'$ among all the truncated patterns.
 We will show that $P\in X$ iff $P$ truncate matches $P_i'\alpha^y$. Since $P_i'$ is a suffix of $S'$ it follows immediately that all patterns that truncate matches $P_i'\alpha^y$ also truncate matches $S'\alpha^y$.
 We will show by contradiction that $P\in X$ implies that 
 $P$ truncate matches $P_i'\alpha^y$.
 Assume that $P=P'\alpha^x\in X$ but does not truncate match $P_i'\alpha^y$. Since $P$ truncate matches $S'\alpha^y$ we have that $x\leq y$ and $P'$ is a suffix of $S'$. Thus if $P'\alpha^x$ does not truncate match $P_i'\alpha^y$ it must be the case that $P'$ is not a suffix of $P_i'$.
 
 But then $P'$ is a longer suffix of $S'$ than $P_i'$ contradicting that $P_i'$ is the longest suffix of $S'$ among all truncated patterns. 
 Thus the patterns that truncate match $P_i'\alpha^y$ is exactly $X$ and by querying the \tmr\ data structure with $(i, \alpha, y)$, we report all occurrences in $X$ and hence the occurrences which end in the unprocessed run.

 Finally, we will show that after step 2 the string $\nodeString{v}$ is the longest suffix of $S'\alpha^y$. Let $w$ be the node set to $v$ by the end of step 2, i.e., after iteration $t$ of the algorithm $v=w$. 
  Assume for the sake of contradiction that after step 2, $\nodeString{w}$ is not the longest suffix of $S'\alpha^y$.
 Then either $\nodeString{w}$ is not a suffix of $S'\alpha^y$ or there is another node $u$ such that $\nodeString{u}$ is a suffix of $S'\alpha^y$ and $\nodeString{w}$ is a proper suffix of $\nodeString{u}$.
 By the definition of the failure links and the dictionary $D_v$ after step 2 $\nodeString{w}$ must be a suffix of $S'\alpha^y$. Furthermore, either $\nodeString{w} = \epsilon$ ($w$ is the root) or $\alpha^y \in D_{p(w)}$, where $p(w)$ is the parent of $w$.
 Assume that there is a node $u$ such that $\nodeString{u}$ is a suffix of $S'\alpha^y$ and $\nodeString{w}$ is a proper suffix of $\nodeString{u}$.
 Then there exist nodes $v'$ and $u'$ such that $\nodeString{w}=\nodeString{w'}\alpha^y$ and $\nodeString{u}=\nodeString{u'}\alpha^y$ and $\alpha^y\in D_{w'}\cap D_{u'}$.
 Since $\nodeString{w}$ is a suffix of $\nodeString{u}$ then $\nodeString{w'}$ is a suffix of $\nodeString{u'}$.
 At the beginning of step 2 $\nodeString{u'}$ must be a suffix of $\nodeString{v}$ since by the induction hypothesis $\nodeString{v}$ is the longest suffix of $S'$.
 Since $u'$ is longer than $w'$ we meet $u'$ before $w'$ in the traversal in step 2. Since $\alpha^y \in D_{u'}$ we would have stopped at node $u'$ and not $w'$.

\subparagraph{Preprocessing}
First we construct the run-length encoded trie $T_\text{RLE}$ and the \tmr\ data structure.
In order to compute the failure link $F_u$ for a non-root node $u$ observe that if the edge $(v,u)$ is labeled $\alpha^x$, then there are 3 scenarios which determine the value of $F_u$.
\begin{itemize}
    \item Either $F_u=D_w[\alpha^x]$ where $w$ is the first node such that $\alpha^x\in D_w$ which is reached by repeatedly following the failure links starting at node $F_v$.
    \item If no such node exists then if the root $r$ has an outgoing edge $\alpha^y$ where $y< x$ then $F_u = D_r[\alpha^{y'}]$ where $y'$ is maximum integer $y'< x$ such that there is an outgoing edge $\alpha^{y'}$ from the root $r$.
    \item Otherwise $F_u = r$.
\end{itemize}
We compute the failure links by traversing the trie in a Breadth-first traversal starting at the root using the above cases.
Note that the second case requires a predecessor query.
In our preprocessing of the $i_v$ values we use the reverse failure links $F'_v$ of node $v\in T_\text{RLE}$, i.e., $u\in F'_v$ if $F_u=v$. We compute these while computing the failure links. 
To compute $i_v$, we first set $i_v$ for each node $v$ which has a child $u$ corresponding to a pattern.
We then start a Breadth-first traversal of the graph induced by the reverse failure links from each of these nodes. 
For each node $v$ we visit, we set $i_v=i_u$ for the parent $u$ of $v$ in the traversal. Note that each node is visited in exactly one of these traversals, since a node has exactly one failure link. 
For the remaining unvisited nodes let $i_v=-1$.

\subparagraph{Preprocessing Time Analysis}
We can construct the run-length encoded trie of the patterns $\patterns$ in expected $O(\overline{m} + k \log{\log{k}})$ time by \cref{cor:rle_sort} and the proof of \cref{lm:rle_sort_blackbox}.
We use expected $O(\overline{m})$ time to construct the dictionaries $D_v$ of the nodes $v$ in the run-length encoded trie $T_\text{RLE}$ and $O(\overline{m} + k\log{\log{k}})$ for the \tmr\ data structure by \cref{lm:ctsr}.
Since a failure link points to a proper suffix, each root-to-leaf path in $T_\text{RLE}$ can traverse a number of failure links proportional to the number of characters on the path. 
The sum of characters of all root-to-leaf paths is $O(m)$; hence, we traverse at most $O(m)$ failure links as we construct the failure links.
We use at most one predecessor query for each node in $T_\text{RLE}$ as we construct the failure links and thus we use $O(\overline{m}\log{\log{m}} + m)$ time to construct the failure links by \cref{lm:predecessor}.
Finally, computing $i_v$ requires $O(\overline{m})$ time to traverse $T_\text{RLE}$.
In total, we use $O(\overline{m}\log{\log{m}} + m + k\log{\log{k}})=O(\overline{m}\log{\log{m}} + m)$ in expectation.

\subparagraph{Dealing with Single Run Patterns}
To correctly report the occurrences of the single run patterns,
we construct a dictionary $D$ such that $D[\alpha]$ is the list of single run patterns of the character $\alpha$ sorted by their length.
Let $S'$ be the processed part of $S$ and let $\alpha^y$ be the next unprocessed run in $S$.
To report the single run patterns occurring in the run $\alpha^y$ we do a linear scan of $D[\alpha]$.
For each single run pattern $\alpha^x$ in $D[\alpha]$ the pattern $\alpha^x$ occurs in all the locations $|S'| + i$ for $0 \leq i\leq y-x$.
If $x>y$ then $\alpha^x$ does not occur in $\alpha^y$ and neither does any pattern later in the list $D[\alpha]$ since they are sorted by length.
We can identify the single run patterns and construct $D$ in $O(\overline{m} + k\log{\log{k}})$ expected time and reporting the occurrences of the single run patterns takes $O(1 + \text{occ})$ time, neither of which impact our algorithm.

In summary, we have shown the following: 

\begin{lemma}
    Let $\mathcal{P} = \{P_1, \ldots P_k\}$ be a set of $k$ strings of total length $m$ and let $S$ be a string of length $n$. Given the RLE representation of $\mathcal{P}$ and $S$ consisting of $\overline{m}$ and $\overline{n}$ runs, respectively, we can solve the dictionary matching problem in $O(\overline{n}\log{\log{k}} + \overline{m}\log{\log{m}} + n + m + \text{occ})$ expected time and $O(\overline{m})$ space, where $\mathrm{occ}$ is the total number of occurrences of $\mathcal{P}$ in $S$. 
\end{lemma}
In the next section, we show how to improve the time bound to $O((\overline{m} + \overline{n})\log{\log{m}} + \text{occ})$ expected time, thus effectively removing the linear dependency on the uncompressed lengths of the string and the patterns.

\section{Full Algorithm} 
\label{sec:full_structure}
In the simple solution the failure links could point to a suffix that might only be one character shorter. Thus navigating them during the algorithm can take $\Omega(n)$ time.
In this section, we show how to modify the failure links such that they point to a suffix that has fewer \emph{runs}.
The main idea is to group nodes that only differ by the length of their first run and navigate them simultaneously.

\subparagraph{Data Structure} We build the same structure as in \cref{sec:static:compressed}, with a slight alteration to the failure links.
Let $v$ be a node in the trie $T_\text{RLE}$ and let $\beta$
be the first character in $\nodeString{v}$, i.e., $\nodeString{v}=\beta^xX$ for some $x$. The failure link $F_v$ stores a pointer to the node $u$ in $T_\text{RLE}$ such that $\nodeString{u}$ is the longest suffix of $X$.  Additionally, we store the length $x$ of the first run of $\nodeString{v}$  as $Z_v=x$. We define the failure link for the root to be the root itself.

\begin{figure}[t]
    \centering
    \includegraphics[scale=1]{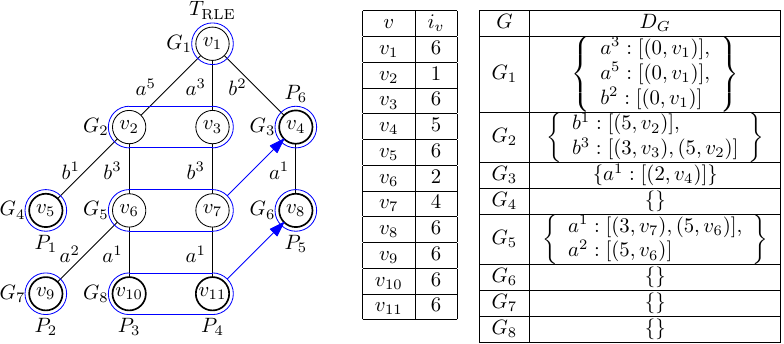}
    
    \caption{
    The structure of the full solution for the 6 patterns $\{a^5b^1, a^5b^3a^2, a^5b^3a^1, a^3b^3a^1, b^2a^1, b^2\}$.
    Here the blue enclosures indicate grouped nodes, and the blue arrows indicate the failure link that all nodes in the group have in common.
    Failure links going to the root node have been left out for simplicity.
    Note that the patterns are the same as in \cref{fig:reporting}.
    }
    \label{fig:advanced}
\end{figure}

We partition the nodes of $T_\text{RLE}$ into groups as follows:
Let $v$ and $u$ be two nodes of $T_\text{RLE}$. If $\nodeString{v}=\beta^xX$ and $\nodeString{u}=\beta^yX$, then $v$ and $u$ belong to the same group $G$.
We define the group of the root to consist only of the root node.
See \cref{fig:advanced}.
For a group $G$ let $G_{\alpha,x}$ be all the vertices $v\in G$ where $v$ has an outgoing edge labeled $\alpha^x$.

For each group $G$, we store the following:
\begin{itemize}
    \item $D_G$: A dictionary of the labels of the outgoing edges of the nodes in $G$. For each label $\alpha^x$, 
     $D_G[\alpha^x]$ is a predecessor structure of the nodes $v\in G_{\alpha,x}$ ordered by the length of their first run $Z_v$.
\end{itemize}
The dictionaries use linear space with the number of outgoing edges of $G$.
Since the number of edges and nodes in $T_\text{RLE}$ is $O(\overline{m})$ and the groups partition the nodes of $T_\text{RLE}$, then the combined size of the dictionaries of the groups is $O(\overline{m})$.
In total, we use $O(\overline{m})$ space.

\subparagraph{Query}
Given a run-length encoded string $S$ we report all locations of occurrences of the patterns $\patterns$ in $S$ as follows.
We process $S$ one run at a time.
Let $S'$ be the processed part of $S$.
We maintain the node $v$ in $T_\text{RLE}$ such that $\nodeString{v}$ is the longest suffix of $S'$.
Initially, $S'=\epsilon$ and $v$ is the root.
Let $\alpha^y$ be the next unprocessed run in $S$, and let $G$ denote the group of $v$.
\begin{itemize}
    \item[] {\bf Step 1: Report Occurrences.\;} Query the \tmr\ data structure with $(i_v, \alpha, y)$ unless $i_v=-1$.
    For each pattern $j$ returned by the structure, report that pattern $j$ occurs at location $|S'| - |P_j| + x_j$.

    \item[] {\bf Step 2: Follow Failure Links.\;} While $v$ is not the root and $Z_v$ does not have a predecessor in $D_G[\alpha^y]$ set $v=F_v$.
    Finally, if $D_G[\alpha^y]$ has a predecessor $u$ to $Z_v$ then $v=D_u[\alpha^y]$.
\end{itemize}

\subparagraph{Time Analysis}
At each node we visit when processing $S$, we use $O(\log{\log{m}})$ time to determine if we traverse an edge in the group or a failure link by \cref{lm:predecessor}.
We traverse one edge per run in $S$ and thus traverse at most $O(\overline{n})$ edges. 
Since the suffix a failure link points to has fewer runs and the number of runs in $S$ is $\overline{n}$, then we traverse at most $O(\overline{n})$ failure links.
Furthermore, we report from at most one node per run in $S$, and by \cref{lm:ctsr} we use $O(\overline{n}\log{\log{k}} + occ)$ time to report the locations of the matched patterns.
In total, we use $O(\overline{n}\log{\log{m}} + occ)$ time.

\subparagraph{Correctness}
By induction on the iteration $t$, we will show that $\nodeString{v}$ is the longest suffix of $S'$ in $T_{\text{RLE}}$ and that we report all starting positions of occurrences of the patterns $\patterns$ in $S'$.
Initially, $t=0$, $S'=\epsilon$, and $v$ is the root of $T_\text{RLE}$ and thus $\nodeString{v}=\epsilon$.
Assume that at the beginning of iteration $t$, $\nodeString{v}$ is the longest suffix of $S'$ in $T_\text{RLE}$.
Let $\alpha^y$ be the next unprocessed run in $S$.
By the same argument as in \cref{sec:static:compressed}, we report all the occurrences of the patterns that end in the unprocessed run $\alpha^y$.
What remains is to show that after step 2 the string $\nodeString{v}$ is the longest suffix of $S'\alpha^y$.
Let $w$ be the node we set to $v$  in the end of step 2, i.e., after iteration $t$ of the algorithm $v=w$.
Assume for the sake of contradiction that after step 2, $\nodeString{w}$ is not the longest suffix of $S'\alpha^y$.
Then either $\nodeString{w}$ is not a suffix of $S'\alpha^y$ or there is another node $u$ such that $\nodeString{u}$ is a suffix of $S'\alpha^y$ and $\nodeString{w}$ is a proper suffix of $\nodeString{u}$.
By the definition of the failure links and the dictionary $D_G$ after step 2 $\nodeString{w}$ must be a suffix of $S'\alpha^y$.
Furthermore, either $\nodeString{w} = \epsilon$ ($w$ is the root) or $\alpha^y \in D_{p(w)}$, where $p(w)$ is the parent of $w$.
Assume that there is a node $u$ such that $\nodeString{u}$ is the longest suffix of $S'\alpha^y$ and $\nodeString{w}$ is a proper suffix of $\nodeString{u}$.
Then there is a node $u'$ such that $\nodeString{u}=\nodeString{u'}\alpha^y$ and $\alpha^y\in D_{u'}$.
At the beginning of step 2, $\nodeString{u'}$ must be a suffix of $\nodeString{v}$ since by the induction hypothesis $\nodeString{v}$ is the longest suffix of $S'$.
Hence in step 2 we would have stopped at a node $v'$ in the group $G'$ of node $u'$.
Furthermore, $Z_{v'} \geq Z_{u'}$ since $\nodeString{u'}$ is a suffix of $\nodeString{v'}$. 
Since $\alpha^y \in D_{u'}$ then $\alpha^y \in D_{G'}$ and in the predecessor structure $D_{G'}[\alpha^y]$ the node $u'$ will have $Z_{u'}$ as a key and since $Z_{v'}\geq Z_{u'}$ and $\nodeString{u}$ is the longest suffix of $S'\alpha^y$ then $u'$ will be the predecessor of $Z_{v'}$ and thus after step 2 $w=D_{u'}[\alpha^y] = u$.

\subparagraph{Preprocessing}
First we construct the run-length encoded trie $T_\text{RLE}$ and the \tmr\ data structure.
We then compute the groups of $T_\text{RLE}$ as follows.
First we group the children of the root $r$ based on the label on the outgoing edge, such that all children $v$ of $r$ with an $\alpha$ on the edge $(r,v)$ are in the same group.
We compute the groups of all descendent nodes as follows.
Given a group $G$ all the nodes in $D_G[\alpha^x]$ constitute a new group $G'$ for $\alpha^x \in D_G$. 
We then compute failure links between groups.
Note that since a failure link points to a suffix with at least one less run,
all the nodes $v$ in a group $G$ will have the same failure link.
We compute the failure link $F_u$ for a non-root node $u$ as follows.
Let the label of edge $(v,u)$ be $\alpha^x$. There are 3 scenarios which determine the value of $F_u$.
\begin{itemize}
\item Either $F_u= D_w[\alpha^x]$ where $w$ is the first non-root node such that $\alpha^x \in D_w$ which is reached by repeatedly following the failure links starting at node $F_v$.
\item if no such node exists and $u$ is not a child of the root $r$ and the root $r$ has an outgoing edge $\alpha^y$ where $y\leq x$ then $F_u=D_r[\alpha^{y'}]$ where $y'$ is the maximum integer $y' \leq x$ such that there is an outgoing edge $\alpha^{y'}$ from the root $r$.
\item Otherwise $F_u=r$.
\end{itemize}
We compute the failure links by traversing the trie in a breadth-first traversal starting at the root using the above cases.
Note that the second case requires a predecessor query.
To compute $i_v$ we will simulate the reverse failure links used in \cref{sec:static:compressed}.
Observe that the failure links of \cref{sec:static:compressed} are composed of failure links where both endpoints are in the same group and failure links where the endpoints are in different groups.
Since a failure link points to a suffix with at least one less run, we have only computed the failure links between groups.
To simulate the failure links within a group observe that if $v,u \in G$ and $\nodeString{v}=\beta^x$ and $\nodeString{u}=\beta^y$ there is a failure link going from $v$ to $u$ in the data structure of \cref{sec:static:compressed} iff $y < x$ and there is no node $w\in G$ where $\nodeString{w} =\beta^z$ and $y < z < x$.
Hence by building a predecessor structure of the nodes in $G$ we can simulate the failure links of \cref{sec:static:compressed} and compute $i_v$.

\subparagraph{Preprocessing Time Analysis}
We can construct the run-length encoded trie of the patterns $\patterns$ in expected $O(\overline{m} + k \log{\log{k}})$ time by \cref{cor:rle_sort} and the proof of \cref{lm:rle_sort_blackbox}.
We use expected $O(\overline{m})$ time to construct the dictionaries $D_v$ of the nodes $v$ in the run-length encoded trie $T_\text{RLE}$ and $O(\overline{m} + k\log{\log{k}})$ for the \tmr\ data structure by \cref{lm:ctsr}.
To partition $T_\text{RLE}$ into groups we use $O(\overline{m})$ time, and $O(\overline{m}\log{\log{m}})$ to construct the predecessor structures of the groups by \cref{lm:predecessor}.
Since a failure link points to a suffix with at least one less run, each root-to-leaf path in $T_\text{RLE}$ can traverse a number of failure links proportional to the number of runs on the path. 
The sum of runs of all root-to-leaf paths is $O(\overline{m})$; hence, we traverse at most $O(\overline{m})$ failure links as we construct the failure links.
We use at most one predecessor query for each node in $T_\text{RLE}$ as we construct the failure links and thus we use $O(\overline{m}\log{\log{m}})$ time to construct the failure links by \cref{lm:predecessor}.
Finally, computing $i_v$ requires $O(\overline{m}\log{\log{m}})$ time to traverse $T_\text{RLE}$ and simulate the failure links.
In total, we use $O(\overline{m}\log{\log{m}} + k\log{\log{k}})=O(\overline{m}\log{\log{m}})$ expected time. In summary, we have shown \cref{thm:rle_reporting}.

\section{Deterministic Solution}
\label{sec:determ}
In this section, we provide a deterministic solution, eventually arriving at \cref{thm:rle_reporting_determ}.

\subparagraph{Deterministic Toolbox}
We need the following deterministic results on sorting and predecessors.
\begin{lemma}[Han~\cite{DBLP:journals/jal/Han04}]\label{lm:sorting}
Given a set $Q$ of $n$ integers from a universe of size $u$, we can deterministically sort $Q$ in $O(n\log{\log{n}})$ time and linear space.
\end{lemma}

Ružić~\cite[Theorem 3]{DBLP:conf/icalp/Ruzic08} showed that we can have deterministic perfect hashing, if we allow the construction time to be $O(n\lg^2\lg{n})$.
By combining the deterministic perfect hashing with the well-known $y$-fast trie of Willard~\cite{DBLP:journals/ipl/Willard83}, we can support predecessor queries with the following bounds.

\begin{lemma}
Given a set of $n$ integers from a universe of size $u$, we can support predecessor queries in $O(n)$ space, $O(n \log \log n)$ preprocessing time, and $O(\log \log u)$ query time. 
\end{lemma}
\begin{proof}
First the set is sorted using \cref{lm:sorting}.
By setting the size of the buckets in the $y$-fast trie to $\log{u}\log^2\log{n}$, the number of elements in the trie is $O(\frac{n\log{u}}{\log{u}\log^2\log{n}})$, hence we have time to compute the deterministic perfect hashing scheme of Ružić~\cite[Theorem 3]{DBLP:conf/icalp/Ruzic08}.
We can navigate to the correct bucket by a binary search on the levels of the $y$-fast trie in $O(\log{\log{u}})$ time, and we can find the correct value in the bucket using binary search in $O(\log(\log{u}\log^2\log{n})) = O(\log{\log{u}})$.
\end{proof}

\subparagraph{Alphabet Reduction}
Now, we will show that we can reduce the alphabet $\sigma$ to $O(\text{min}(\overline{n}, \overline{m}) + 1)$ by a rank reduction.
\begin{corollary}\label{cor:rank_reduction}
    Let $\mathcal{P} = \{P_1, \ldots, P_k\}$ be a set of $k$ strings of total length $m$ and let $S$ be a string of length $n$ from an alphabet of size $\sigma$.
     Given the RLE representation $\overline{\mathcal{P}}=\{\overline{P_1}, \ldots, \overline{P_k}\}$ of $\mathcal{P}$ and $\overline{S}$ of $S$ consisting of $\overline{m}$ and $\overline{n}$ runs, respectively,
     we can in $O((\overline{n}+\overline{m})\log{\log{(\overline{n} + \overline{m}})})$ time construct the RLE representation $\overline{\mathcal{P}}' = \{\overline{P_1}', \ldots, \overline{P_k}'\}$ of $\mathcal{P'}$ and $\overline{S}'$ of $S'$, such that $\mathcal{P}'$ and $S'$ are from an alphabet $\sigma'=O(\text{min}(\overline{n}, \overline{m}) + 1)$ and iff $S[i] = P_j[z]$ then $S'[i] = P_j'[z]$ for $1 \leq i \leq n, 1 \leq j \leq k$, and $1\leq z \leq |P_j|$.
\end{corollary}
\begin{proof}
    We sort the characters of all the runs in $\overline{\mathcal{P}}$ and $\overline{S}$.
    We then identify all characters in $S$ that do not appear in any pattern by a linear scan of the sorted sequence and replace all of these characters in each run in $\overline{S}$ with the character $\alpha_1$.
    We then identify all characters in the patterns that do not appear in $S$ and replace all of these characters in each run in $\overline{\mathcal{P}}$ with the character $\alpha_2$.
    We then replace the remaining characters in each run of $\overline{\mathcal{P}}$ and $\overline{S}$ with $\alpha_{r+2}$ where $r$ is the corresponding rank in the sorted sequence, disregarding duplicates and characters that only appear in $S$ or only appear in the patterns $\mathcal{P}$.
\end{proof}

We can now, for the remainder of the section, assume that $\sigma = O(\text{min}(\overline{n}, \overline{m}) + 1) = O(\overline{m})$.
This greatly simplifies the construction of the compact trie, since we can now use a single direct addressing table in \cref{cor:rle_sort} and thereby achieve \cref{cor:rle_compact_trie} deterministically with the same complexity (after the strings have been rank reduced).

\subparagraph{Truncate Match Reporting}
In our truncate match reporting data structure, we use randomization in the following places:
\begin{enumerate}
    \item \label{ite:1} The first color ancestor data structure.
    \item \label{ite:2} Construction of the induced trees in \cref{lm:catr}.
    \item \label{ite:3} The dictionary associating each color in $c \in C_v$ for $v\in T$ with the corresponding node in the induced tree $T_c$.
    \item \label{ite:4} The weight functions $\pi_v$.
\end{enumerate}

We can resolve \ref{ite:3} and \ref{ite:4} by using a predecessor data structure in place of a dictionary and use the elements rank to represent the weight function with a direct address table.
By observing that the Euler ordering in the first color ancestor data structure of Muthukrishnan and Müller \cite{DBLP:conf/soda/MuthukrishnanM96} are sorted, we can improve their preprocessing time to become deterministic by replacing the randomized predecessor structure (\cite[Lemma 2.2]{DBLP:conf/soda/MuthukrishnanM96}) with the deterministic predecessor structure of \cref{lm:predecessor}, and hence resolve \ref{ite:1}.
Finally, we can use a direct address table in the construction of the induced trees and thereby resolve \ref{ite:2}.

We use $O(k\log{\log{k}})$ time to sort $C_v$ for $v\in T$ and construct their predecessor data structures and weight functions.
Since the Euler ordering is already sorted, we do not use any additional time on the construction of the first color ancestor data structure.
We use $O(\sigma) = O(\overline{m})$ space during the construction of the induced trees.

To answer a truncate match reporting query, we now also have to perform a predecessor query to access $C_v$, and hence we use an additional $O(\log{\log{\sigma}}) = O(\log{\log{\overline{m}}})$ time.

\subparagraph{Deterministic Run-Length Encoded Dictionary Matching}
Recall that the alphabet size is $\sigma=O(\min(\overline{n}, \overline{m}) + 1)$ due to the rank reduction.
We sort the strings and construct the compact trie with \cref{cor:rle_compact_trie} using a single direct address table instead of a dictionary.
We replace the dictionaries in the run-length encoded trie $T_\text{RLE}$ with predecessor search data structures.
Finally, we use the deterministic truncate match reporting data structure.
The remainder of the construction and preprocessing is equivalent.

The rank reduction requires $O((\overline{n}+\overline{m})\log{\log{(\overline{n} + \overline{m}})})$ time and $O(\overline{n} + \overline{m})$ space.
The compact trie $T_\text{RLE}$, truncate match reporting structure, and the dictionaries in the run-length encoded trie can all be constructed in $O(\overline{m} + k\log{\log{k}})$ time and $O(\overline{m} + k)$ space.
Finally, in the preprocessing step of the full solution in \cref{sec:full_structure}, we query the dictionaries of $T_\text{RLE}$ $O(\overline{m})$ times, hence we use $O(\overline{m}\log{\log{\overline{m}}})$ additional time in the preprocessing step. 
Finally, when we parse $S$ we use $O(\log{\log(\overline{m} + k)}) = O(\log{\log{\overline{m}}})$ time per query to the truncate match reporting structure and $O(\log{\log{\overline{m}}})$ time per query to the dictionaries of the run-length encoded trie $T_\text{RLE}$.
Thus, in total, we use $O(\overline{n}\log{\log{m}} + \text{occ})$ time to report the locations of the matched patterns.
In summary, we have shown \cref{thm:rle_reporting_determ}.

\bibliography{bibliography}

\end{document}